\documentclass[particles,communication,submit,moreauthors,pdftex,10pt,a4paper]{Definitions/mdpi} 
\preto{\abstractkeywords}{\nolinenumbers}

\usepackage{mathtools}

\firstpage{1} 
\pubvolume{xx}
\issuenum{1}
\articlenumber{5}
\pubyear{2018}
\copyrightyear{2018}
\history{Received: date; Accepted: date; Published: date}
\Title{Low-momentum pion enhancement from schematic hadronization of a gluon-saturated initial state}

\Author{Elizaveta Nazarova$^{1,2}$\orcidA{}, \L{}ukasz Juchnowski$^{2}$, David Blaschke$^{2,3,4,*}$ and Tobias Fischer$^{2}$}

\AuthorNames{Elizaveta Nazarova, Lukasz Juchnowski, David B. Blaschke, Tobias Fischer}

\address{%
$^{1}$ \quad Skobeltsyn Institute of Nuclear Physics, Lomonosov Moscow State University, RU-119991 Moscow, Russia\\
$^{2}$ \quad Institute of Theoretical Physics, University of Wroc{\l}aw, 50-204 Wroc{\l}aw, Poland\\
$^{3}$ \quad Bogoliubov Laboratory for Theoretical Physics, JINR Dubna, 141980 Dubna, Russia\\
$^{4}$ \quad National Research Nuclear University (MEPhI), 115409 Moscow, Russia}

\corres{Correspondence: david.blaschke@gmail.com}

\abstract{%
    We study the particle production in the early stage of the ultrarelativistic heavy-ion collisions.
    To this end the Boltzmann kinetic equations for gluons and pions with elastic rescattering are considered together with a simple model for the parton-hadron conversion process (hadronisation). 
    It is shown that the overpopulation of the gluon phase space in the initial state leads to an intermediate stage of Bose enhancement in the low-momentum gluon sector which due to the gluon-pion conversion process is then reflected in the final distribution function of pions. This pattern is very similar to the experimental finding of a low-momentum pion enhancement in the ALICE experiment at CERN LHC.
    Relations to the thermal statistical model of hadron production and the phenomenon of thermal and chemical freeze-out are discussed in this context.%
}


\begin{document}

\section{Introduction}

One of  the issues  which can be addressed by the kinetic approach is the question of a low-momentum  pion enhancement in heavy ion collisions \cite{Abelev:2013pqa}.
%\footnote{The low-transverse-momentum pion spectra show up to 50 \% enhancement compared to hydrodynamic models 
%J\cite{Abelev:2013pqa,JuchnowskiPhd}}.
There are several solutions proposed to explain this effect as, e.g.,  the hadronization and freeze-out in a chemical non-equilibrium \cite{Kataja:1990tp,Gavin:1991ki,Begun:2014aha}, the separate freeze-out for strange particles \cite{Begun:2014rsa}, 
%an incomplete list of hadrons in the resonance gas \cite{Begun:2013nga}, 
Bose-Einstein condensate of pions \cite{Begun:2013nga,Blaizot:2011xf,Begun:2015yco,Begun:2015ifa,Voskresensky:1996ur}, established by elastic rescattering in the final stage \cite{Voskresensky:1996ur,Semikoz:1995rd}. 
However, none of them is commonly accepted yet \cite{Begun:2015yco}.
We believe, an explanation linked to the presence of non-equilibrium physics and a precursor of pion condensation in heavy ion collisions
should be the favorable one, especially after the recent analysis of particle correlations performed  by the ALICE collaboration is showing a coherent fraction of charged $\pi-$meson emission that is reaching 23\% \cite{Abelev:2013pqa,Begun:2015ifa}. 
%Such  Bose condensate formation is usually  model by  introduction of additional non-equilibrium parameters to standard statistical approach.  \cite{Koch:1985hk,Begun:2015yco}. 
Such formation of Bose condensate is usually described by the introduction of additional non-equilibrium parameters to the statistical approach 
\cite{Voskresensky:1995tx,Voskresensky:1996ur}, see also \cite{Kataja:1990tp,Begun:2015yco,Koch:1985hk}. 

An alternative scheme may rely on the Boltzmann kinetic equation for gluons and pions with elastic rescattering and a simple model for the 
parton-hadron conversion process (hadronisation).
There are deep physical reasons for the non-equilibrium and pion condensation at the LHC. 
It can be due to fast expansion and overcooling of the QGP, or due to gluon condensation
in the color glass condensate (CGC) initial state preceeding subsequent hadronization of the low-momentum gluons into low-momentum 
pions \cite{Begun:2015yco}. 
A scenario with an initial state dominated by gluons which subsequently hadronize, eventually via a quarkless evolution through a
first order phase transition, has recently been considered in Ref.~\cite{Stoecker:2015zea}.

In this short communication we investigate the idea that a certain oversaturation of the purely gluonic initial state could lead by elastic rescattering to a precursor of Bose condensation in the gluon sector in the form of a low-momentum gluon enhancement which, however, should be depopulated by the gluon-pion conversion process and thus appear as low-$p$ pion enhancement in the pion sector. 
The gluon-pion conversion process is assumed with a constant matrix element which may be pictured as the local limit of a quark one-loop diagram for the case of large quark mass (quark confinement). 
We demonstrate the evolution of the coupled gluon and pion distribution functions in this case within a schematic model of coupled kinetic equations.

%The non-equilibrium hadronization can explain the measured particle ratios [8], and also
%the spectra [16, 17] very well. The obtained parameters indicate the possibility that 5% of the total
%number of pions can be in BEC at the LHC [18]. 

%We should focus in this paper on the introduction of the system, the idea 
%of the investigation (precursor of Bose condensation in gluon sector gets
%mapped to the low-p pion enhancement by the conversion process).
%For this we show the numerical results on the distribution functions and
%discuss them and that's it. 

\section{Kinetic equation approach to thermalization and hadronization}
\label{main}

We start with the kinetic equation in the form of a Boltzmann-Nordheim equation, which for a single particle distribution function $f=f(\vec{x},\vec{p},t)$ can be written as
\begin{eqnarray}
\label{eqn:kinetic_general}
    \frac{\mathrm df}{\mathrm dt} = C[f]\,,
\end{eqnarray}
where 
\begin{eqnarray}
\label{eqn:derivative1}
   \frac{\mathrm df}{\mathrm dt} =	\frac{\partial f}{\partial\vec x}\frac{\mathrm d\vec x}{\mathrm dt}  +\frac{\partial f}{\partial\vec p}\frac{\mathrm d\vec p}{\mathrm dt} + \frac{\partial f}{\partial t}
\end{eqnarray}
and $C[f]$ represents the collision integral. In this study we restrict ourselves to the case of a uniform ($\partial f/ \partial\vec x = 0$) system in a non-expanding box ($\vec F = \mathrm d\vec p/\mathrm d t = 0$), therefore only the explicit time-dependence remains: $\mathrm df/ \mathrm dt = \partial f/ \partial t \equiv \partial_t f$. 

On the other hand, the collision integral for the $1+2 \rightarrow 3+4$ process is defined as:
\begin{eqnarray}\label{eqn:collision_integral1}
	C[f(t,\vec{p}_1)] = \frac{(2\pi)^4}{2\mathrm E_1} \int \delta^4 (\sum_{i} P_{i}) | \mathrm M |^2 F[f] \prod_{k=2}^{4} \frac{\mathrm d^3 \vec{p}_k}{(2\pi)^3 2E_k}~,
\end{eqnarray}
so that the Eq. \eqref{eqn:kinetic_general} will take the following form:
\begin{eqnarray}\label{eqn:equation_full}
	\partial_t f (t,\vec{p}_1) = \frac{(2\pi)^4}{2\mathrm E_1} \int \delta^4 (\sum_{i} P_{i}) | \mathrm M |^2 F[f] \prod_{k=2}^{4} \frac{\mathrm d^3 \vec{p}_k}{(2\pi)^3 2E_k}~,
\end{eqnarray}
describing elastic scattering of the system of particles of one type, e.g. gluons. Here for the process  $1+2 \rightarrow 3+4$ we define as $f_i$  the distribution function of particle $i$ with 4-momentum $P_i = (E_i, \vec{p}_i)$, $|M|$ as the transition amplitude of the process, and $F[f]  = (1 + f_1) (1 + f_2) f_3 f_4 - f_1 f_2 (1 + f_3) (1 + f_4)$ represents the gain and loss terms in the collision integral.
In the current study we consider the distribution function to be isotropic through the whole evolution. Moreover, the matrix elements of all the processes involved are taken to be constant:
\begin{eqnarray}
    \mathrm |M|_{12\rightarrow34} = \mathrm{const}\,,
\end{eqnarray}
following \cite{Semikoz:1995rd}, where the case of a system of pions was considered. Albeit this work describes the academic study with only constant matrix elements, the ongoing project involving momentum- and angle-dependent transition amplitudes is discussed in the section \ref{Discussion}.  

As we consider an isotropic, uniform, non-expanding system and constant matrix elements for the processes, and taking into account the 4-momentum conservation ($P_1 + P_2 = P_3 + P_4$), the equation \eqref{eqn:equation_full} takes the form
\begin{align}\label{eqn:integ_energy}
	\partial_t f (\varepsilon_1) &= \frac{\left|M\right|^2}{64\pi^3 \varepsilon_1} \int \int \mathrm d \varepsilon_3 \mathrm d \varepsilon_4 
	\mathrm D F[f]\,,
\end{align}
where $\mathrm D = \min \{p_1, p_2, p_3, p_4\}$ and $p_i$ are now the radial components of the three-momenta.
Details of the derivation are shown in the Appendix \ref{Appendix1}.
For future investigations it is helpful to rewrite the equation \eqref{eqn:integ_energy} in terms of a momentum integration, as we would like to extend the approach to the angle-dependent collision integral, as well as to non-uniform systems.
Therefore, in the current work we use the following formula:
\begin{align}\label{eqn:integ_momenta}
	\partial_t f (p_1) &= \frac{\left|M\right|^2}{64\pi^3 \varepsilon_1} \int \int \frac{p_3 p_4}{\varepsilon_3 \varepsilon_4} \mathrm d p_3 \mathrm d p_4 
	\mathrm D F[f]\,.
\end{align}

Obviously, elastic scattering is necessary but not sufficient to achieve low-$p$ pion enhancement. 
The second required process which needs to be accounted for is hadronization. 
In this exploratory work we connect the gluon sector directly with the pion one. 
For such system  there are three contributing channels: 
$\pi \pi \rightarrow \pi \pi$, $g g \rightarrow g g$, and $g g \leftrightarrow \pi \pi$. 
Therefore at the end we have a coupled system of equations:

\begin{subequations}
\label{eq:csys}
\begin{align}
\begin{split}
    \frac{\partial f_\pi}{\partial t}(t,\vec{p}_1)
    &=  \int \int\frac{\left|M_{\pi \pi\rightarrow \pi \pi}\right|^2}{64\pi^3 \varepsilon_1} \frac{p_3 p_4}{\varepsilon_3 \varepsilon_4} \mathrm d p_3 \mathrm d p_4 \mathrm D F[f_\pi]\\
    &+ \left(1+f_\pi(t,p_1)\right) \int \int \frac{\left|M_{g g\rightarrow \pi \pi}\right|^2}{64\pi^3 \varepsilon_1} \frac{p_3 p_4}{\varepsilon_3 \varepsilon_4}  \mathrm d p_3 \mathrm d p_4 \mathrm{D }
    \left(1+f_\pi(t,p_2)\right)f_g(t,p_3)f_g(t,p_4) \\
    &- f_\pi(t,p_1) \int \int \frac{\left|M_{\pi \pi\rightarrow g g}\right|^2}{64\pi^3 \varepsilon_1} \frac{p_3 p_4}{\varepsilon_3 \varepsilon_4} \mathrm d p_3 \mathrm d p_4 \mathrm{D } f_\pi(t,p_2)\left(1+f_g(t,p_3)\right)\left(1+f_g(t,p_4)\right)
    \label{eq:csys1}
\end{split}
\end{align}
% % % % % % % % % % % % % % % % % % % % % % % % % % % % % % % % % % % % %

\begin{align}
\begin{split}
    \frac{\partial f_g}{\partial t}(t,\vec{p}_1)
    &= \int \int\frac{\left|M_{g g\rightarrow g g}\right|^2}{64\pi^3 \varepsilon_1} \frac{p_3 p_4}{\varepsilon_3 \varepsilon_4} \mathrm d p_3 \mathrm d p_4 \mathrm D F[f_g]\\
    &+ \left(1+f_g(t,p_1)\right) \int \int \frac{\left|M_{\pi \pi\rightarrow g g}\right|^2}{64\pi^3 \varepsilon_1} \frac{p_3 p_4}{\varepsilon_3 \varepsilon_4}  \mathrm d p_3 \mathrm d p_4 \mathrm{D} \left(1+f_g(t,p_2)\right)f_\pi(t,p_3)f_\pi(t,p_4)\\
    &- f_g(t,p_1) \int \int \frac{\left|M_{g g\rightarrow \pi \pi}\right|^2}{64\pi^3 \varepsilon_1} \frac{p_3 p_4}{\varepsilon_3 \varepsilon_4} \mathrm d p_3 \mathrm d p_4 \mathrm{D} f_g(t,p_2)\left(1+f_\pi(t,p_3)\right)\left(1+f_\pi(t,p_4)\right)
\label{eq:csys2}
\end{split}
\end{align}
\end{subequations}
where $\mathrm M_{g g\rightarrow \pi \pi}$ and $\mathrm M_{\pi \pi\rightarrow g g}$ are matrices for hadronization channels. Note, that due to the momentum conservation $p_2 = p_3 + p_4 - p_1$ in Eq. \eqref{eq:csys}. In this study we set $\mathrm M_{\pi \pi\rightarrow g g} = 0$, which is motivated by the threshold for this process due to the large value of the gluon mass: $m_g = 0.7$~GeV. The value of $\mathrm M_{g g\rightarrow \pi \pi}$ is set to be constant and should be seen as an academic example.

As the initial condition of the system we take an oversaturated gluon distribution given by a step-like function \cite{Blaizot:2013lga,Blaizot:2015wga} inspired by the CGC picture of the initial state which is assumed to have no pions
\begin{align}\label{eq:init}
    f_\pi( t,p )\bigg |_{t=0} &= 0\,,
    & f_g( t,p )\bigg |_{t=0} &= f_0 \, \theta(1 - p/Q_s )\,.
\end{align}
By $Q_s$ we denote the saturation scale. 
However, in  order to avoid numerical problems that  would  occur  with  the  step-function  distribution,  we  use  instead  the following smooth function \cite{Blaizot:2013lga}
\begin{eqnarray}\label{finitCGC}
    f_g(t,p)\bigg |_{t=0} = f_0 \, \left[ \theta(1 - p/Q_s ) + \theta(p/Q_s-1) e^{-a\, (p/Q_s-1)^2} \right], \quad a = 10
\end{eqnarray}
to define the initial conditions.

%\begin{eqnarray}
%&&f_\pi( t,p )\bigg |_{t=t0}
%=  0
%\\
%&&f_g( t,p )\bigg |_{t=t0}
%=  f_0 \, \theta(1 - p/Q_s )
%\label{cos}
%\end{eqnarray}
 
%%%%%%%%%%% hadronization process discussion %%%%%%%%%%%

We keep our model simple and therefore do not introduce an extra timescale for the start of hadronization. 
However, we keep in mind that the underlying microphysical process is, e.g., a quark-box diagram, which consists of the 
Breit-Wheeler type process of $2g \to q \bar{q}$ and subsequent hadronization cross section  $q \bar{q} \to \pi \pi$ .
In the future we plan to investigate the problem of the gluon-to-pion conversion in detail, for instance within a 
Nambu--Jona-Lasinio model \cite{Rehberg:1998me,Rehberg:1995kh,Friesen:2013bta,Marty:2015iwa} and/or by exploiting 
dynamical schemes of hadronization that would address the confinement aspect as well 
\cite{Feng:2016ddr,Florkowski:2015cba,Florkowski:2015rua,Florkowski:2015rxg}.

%%%%%%%%%%%%%%%%%%%%%%%%%%%%%%%%%%%%%%%%%%
\section{Results}
\label{Results}
 
\begin{figure}[tbh]
\centering
	\includegraphics[scale=0.5]{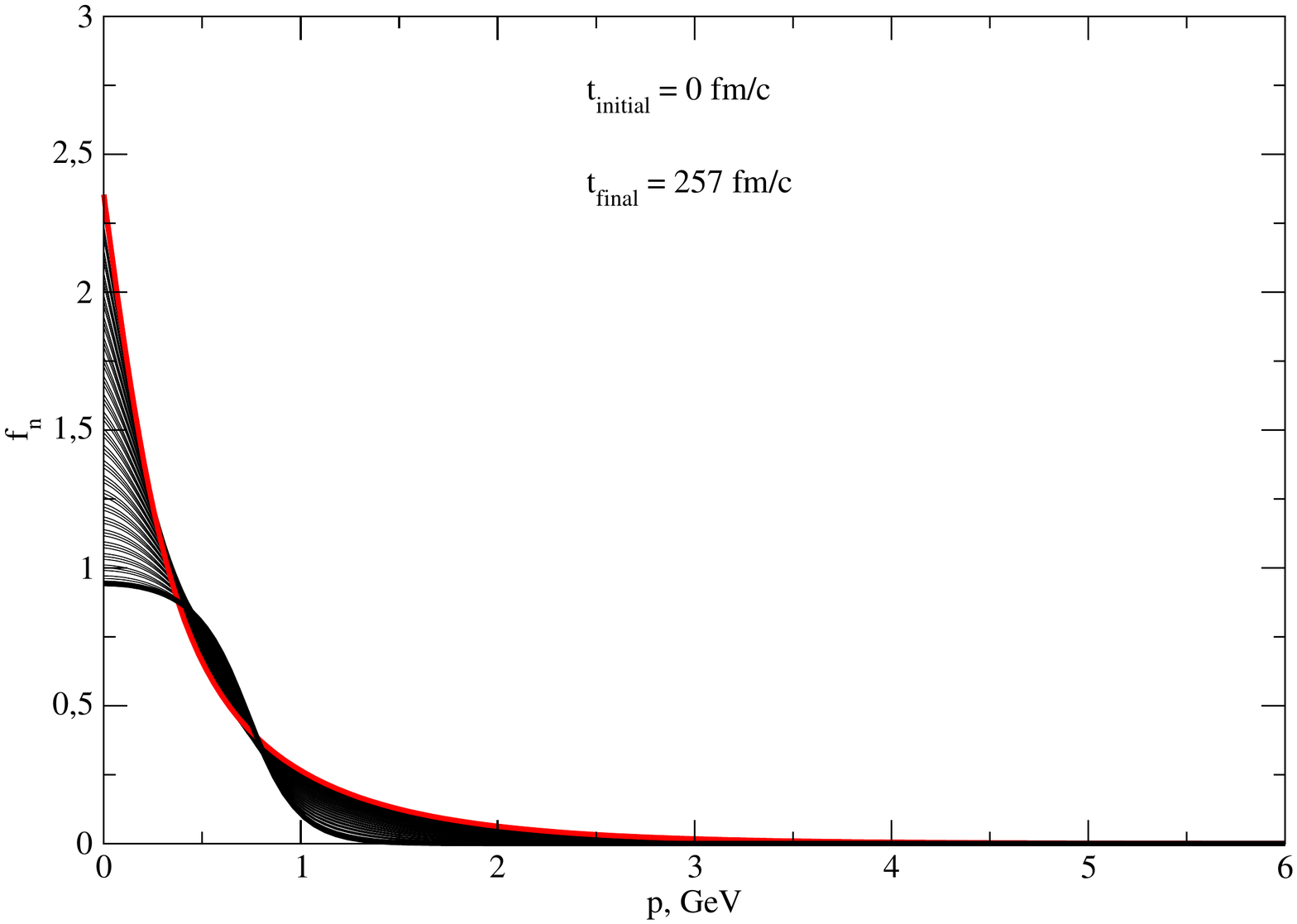}
%\vspace{-10mm}
	\caption{(Color online) The evolution of gluon distribution function $f(p)$ with time in a system of massive
gluons ($m=0.7$~GeV). The final distribution is shown as a bold red line, while the initial function is drawn as a bold black line. Thin black lines represent the intermediate stages of the gluon distribution function. The final time of the evolution represents the point when the pion distribution reaches equilibrium.}
	\label{fig:gluon_evolution}
\end{figure}

 In Fig.~\ref{fig:gluon_evolution} we show the evolution of the gluon distribution function from a CGC motivated initial (over-)saturated gluon state to a thermal distribution due to elastic scattering according to the $gg \to gg$ process. 
The timescale to reach a thermalized final state is of the order of $t_{\rm final} \sim 250 $ fm/c and thus exceeds the typical duration evolution towards freeze-out of the fireball created in a heavy-ion collision. 
This is mainly due to the fact that the value of the matrix element taken in this example calculation as $|M|=4.5$ is unrealistically small. 

\begin{figure}[tbh]
\centering
	\includegraphics[scale=0.2]{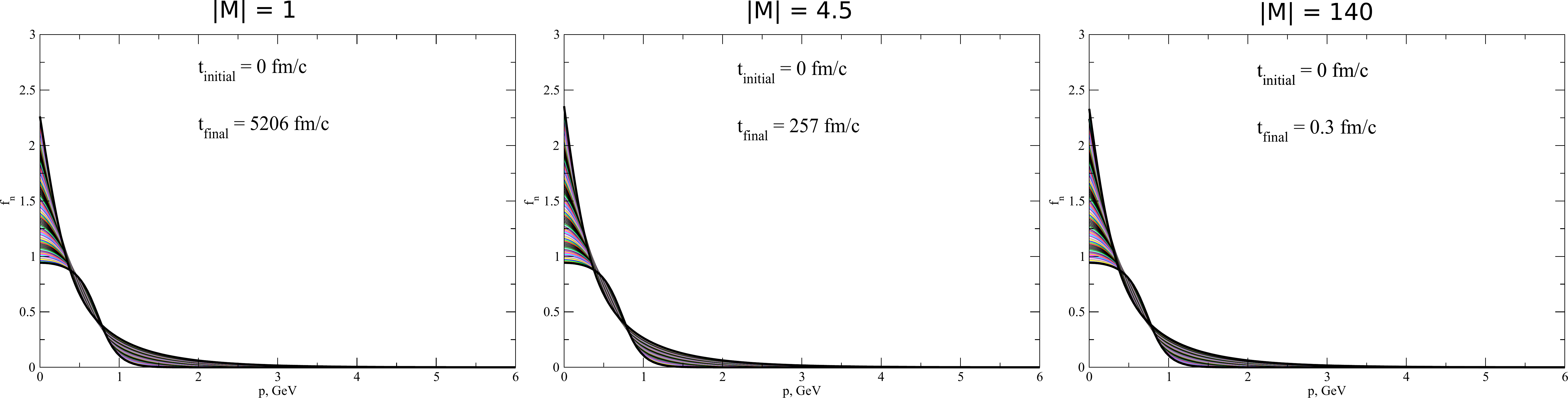}
%\vspace{-10mm}
	\caption{(Color online) Same as Fig.~\ref{fig:gluon_evolution} for different matrix elements $|M|=1$, 4.5, 140.}
	\label{fig:gluon_many}
\end{figure}

In Fig.~\ref{fig:gluon_many} we show the same evolution of the gluon distribution function for three different values of the matrix element.
The value $M=140$ leads to a thermalization time scale which nicely corresponds to the result of a calculation by Shuryak 
\cite{Shuryak:1992wc}.

\begin{figure}[tbh]
    \centering
	\includegraphics[scale=0.5]{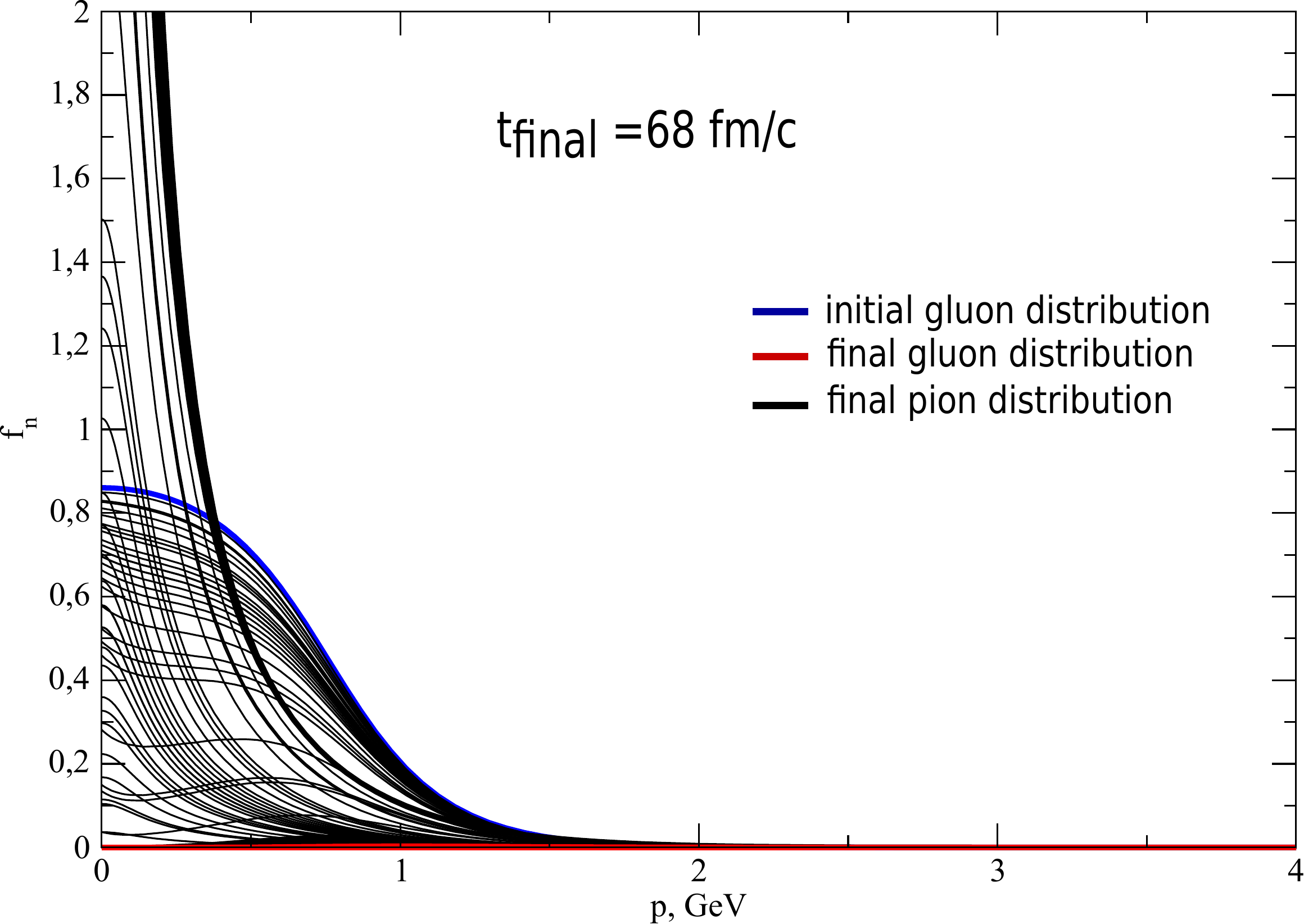}
%\vspace{-10mm}
	\caption{(Color online) The evolution of pion ($m=0.14$~GeV) and gluon ($m=0.7$~GeV) distribution functions $f(p)$ with time in a coupled pion-gluon system. Blue line represents the  initial gluon distribution, while the final distributions are shown as bold black and red lines for pion and gluon distribution functions, respectively.  Thin black lines represent the intermediate stages of the distribution functions. The final time of the evolution represents the point when the pion distribution reaches equilibrium.}
	\label{fig:gluon_pion_evolution}
\end{figure}

When the coupling to the pion sector is switched on, the gluon conversion proceeds and the initially empty pion phase space gets populated at the expense of the gluon one. Due to the relation of the gluon and pion masses the reverse process (the pion annihilation to two gluons) does practically  not take place.
In Fig.~\ref{fig:gluon_pion_evolution} the evolution from the initially pure gluon saturated state to the thermal pion state without gluons is shown.
The pion distribution shows clearly the low-momentum enhancement typical for a precursor of Bose condensation. 
This is the fact observed in the ALICE experiment at CERN for which we wanted to give a qualitative explanation with the simple kinetic model presented here. 
It should be noted that here we used as a test the equal values for the three transition amplitudes: 
$|M_{g g\rightarrow g g}| = |M_{\pi \pi\rightarrow \pi \pi}| = |M_{g g\rightarrow \pi \pi}| = 4.5$. 

%\todo{Figure with evolution of the coupled system:}
 
Our simplified model shows,  under the assumption of gluon dominance in the initial state, the quarkless evolution of the system towards a pion gas with low-momentum pion enhancement as a precursor of Bose condensation. 
According to \eqref{eq:csys} both particle species (gluons and pions) undergo two main processes: conversion and elastic scattering. Both of them are responsible for low-momentum (low-$p$) pion enhancement.

The fist process turns $\pi$-mesons to gluons and vice versa and its rate is defined by two matrix elements $M_{\pi \pi\rightarrow g g}, ~M_{g g\rightarrow \pi \pi}$, which in the simplest case considered here are constant numbers. Particle conversion can take place only when energy of incoming particles is at least equal to mass of outgoing ones. Consequently, in case of massless gluons kinematics restricts  $gg \rightarrow \pi\pi$ reaction  to higher energy region of a spectrum, making the whole process slow.

The impact of the second process  (elastic scattering ) is more subtle. 
It lowers momentum of particles through subsequent collisions, leading them to "pile-up" near zero momentum mode. 
The effect is especially strong for bosons due to the statistical factor $(1+f_{i})$ in \eqref{eq:csys} and allows pre-condensate formation even before thermalization. 
In normal circumstances for  long enough times the distribution should  become an equilibrium  Bose function. 
However, in our model massive  gluons undergo a complete conversion to pions before thermalization because 
$m_g > m_\pi$ ($m_g = 0.7$~GeV, while $m_\pi = 0.14$~GeV). 

%The objective is explanatory one

%There are deep physical reasons for the non-equilibrium and pion condensation at the LHC. It
%can be due to fast expansion and overcooling of the QGP [12, 13], or due to gluon condensation
%in Color Glass Condensate [14], and subsequent hadronization of the low p T gluons into low p T
%pions [15]. The non-equilibrium hadronization can explain the measured particle ratios [8], and also
%the spectra [16, 17] very well. The obtained parameters indicate the possibility that 5% of the total
%number of pions can be in BEC at the LHC [18].

%The second process (elastic scattering )

%Such kinematics restricts $qq \rightarrow \pi\pi$ reaction to high energy region of massless gluon spectrum.
%so massless gluons  

%Particle conversion is necessary but  not sufficient to achieve  pion enhancement.

%\todo{Figure with evolution of the pure gluon system with massless gluons (for comparison with Blaizot):}

%\todo{Maybe a comparison between the results of massless gluon system with constant matrix element and radial-dependent?}

%%%%%%%%%%%%%%%%%%%%%%%%%%%%%%%%%%%%%%%%%%
\section{Discussion}\label{Discussion}

The present model, albeit quite simple, shows the formation of the pion condensation precursor emerging from an oversaturated purely gluonic state. The process takes place before the system reaches equilibrium. 
%Moreover the enhancement of low momentum modes is high enough to dismiss influence of a particle number non-conservation. 
%related to numerical errors.
The model can be improved by the use of non-constant matrix elements and thus taking into account  scattering angle in collision kinematics.
Such an improved model would allow us to discuss the different scales and their evolution, e.g., the Debye scale, the UV and IR scale, see Refs. \cite{Semikoz:1995rd,Blaizot:2011xf,Blaizot:2013lga,Zhou:2017zql}. 

These improved matrix elements should also bear the confining aspects of gluon-gluon interactions which ultimately should be responsible for the absence of gluons from the final state. The assumption of a constant gluon mass, exceeding the value of the pion mass is a rather schematic realization of this concept which provides ample room for improvement. 
Here it would be beneficial to make a comparison with the study in Ref. \cite{Blaizot:2013lga}, where a system of massless gluons undergoes the evolution due to elastic scattering with similar restrictions as used in the current paper. However, the equation \eqref{eqn:integ_momenta} will no longer be valid in the case of non-constant matrix elements and angle-dependence, and thus will need to be rederived. 
%Another space for advancement he lies

Another room for advancement lies in direct handling of the kinetics of Bose condensation (see, e.g., Ref.~\cite{Voskresensky:1996ur}).
One way to do that is the separation of the distribution function into two parts:
\begin{equation}
\tilde{f}_{\pi}(p) = f_{\pi}(p) + (2\pi)^3 n^{\pi}_c \delta(p)
\end{equation}
\begin{equation}
\tilde{f}_{g}(p) = f_{g}(p) + (2\pi)^3 n^{g}_c \delta(p)
\end{equation}
where the first term represents the "gas" and the second describes BEC. 
This ansatz has been discussed for the oversaturated pion gas in Ref. \cite{Semikoz:1995rd} 
and recently also for the gluon plasma in Ref.~\cite{Xu:2014ega}.
We hope to achieve manifest energy and particle number conservation with such an improved formulation of the particle kinetics in the presence or precursory development of a Bose condensate in the system.

The model can be extended towards a more realistic description of a hadronizing gluon-dominated initial state in high-energy heavy-ion collisions by including more hadronic species as they are observed in those experiments in good agreement with the thermal statistical model \cite{Andronic:2017pug}.
This calls then for an extension of the collision integrals in our kinetic model to other classes of processes than just $2\to 2$ 
processes as, e.g., the three-meson conversion to a baryon-antibaryon pair and its reverse \cite{Seifert:2017oyb}.

Last, but not least we want to mention that the assumed absence of dynamical quarks is only a simplifying assumption. 
In an improved model, their kinetics shall be coupled to that of the gluons and all considered hadron species. 
Their absence in the final state shall be realised due to a confining mechanism. The one already tested in the framework of a kinetic theory is the Gribov-Zwanziger confinement realized by an infrared-divergent selfenergy \cite{Florkowski:2015cba,Florkowski:2015rua,Florkowski:2015rxg}.
We shall come back to these issues in a subsequent, more elaborate work on the subject.
%%%%%%%%%%%%%%%%%%%%%%%%%%%%%%%%%%%%%%%%%%
%\section{Conclusions}

%%%%%%%%%%%%%%%%%%%%%%%%%%%%%%%%%%%%%%%%%%
%\authorcontributions{For research articles with several authors, a short paragraph specifying their individual contributions must be provided. The following statements should be used “conceptualization, X.X. and Y.Y.; methodology, X.X.; software, X.X.; validation, X.X., Y.Y. and Z.Z.; formal analysis, X.X.; investigation, X.X.; resources, X.X.; data curation, X.X.; writing—original draft preparation, X.X.; writing—review and editing, X.X.; visualization, X.X.; supervision, X.X.; project administration, X.X.; funding acquisition, Y.Y.”, please turn to the  \href{http://img.mdpi.org/data/contributor-role-instruction.pdf}{CRediT taxonomy} for the term explanation. Authorship must be limited to those who have contributed substantially to the work reported.}

%%%%%%%%%%%%%%%%%%%%%%%%%%%%%%%%%%%%%%%%%%
\funding{
This research was funded by the Polish Narodowe Centrum Nauki under grant number UMO-2014/15/B/ST2/03752 (E.N., L.J., D.B.) and grant number UMO-2016/23/B/ST2/00720 (E.N.,T.F.).
}

%%%%%%%%%%%%%%%%%%%%%%%%%%%%%%%%%%%%%%%%%%
\acknowledgments{
We acknowledge discussions with Viktor Begun, Marcus Bleicher, Wojciech Florkowski, Carsten Greiner, Gerd R\"opke, Ludwik Turko and Dmitry Voskresensky on the topic of the nonequilibrium nature of the pion distribution emerging from an oversaturated initial state.
Further we thank Niels-Uwe Friedrich Bastian for his support in numerical and algebraic questions.
Brent Harrison gave a nice presentation of his ongoing work with Andre Peshier on ``Bose-Einstein Condensation from a Gluon Transport Equation'' at the Symposium on ``Nonequilibrium Phenomena in Strongly Correlated Systems'' in Dubna, 16. April 2018, which inspired us to prepare this communication.%?????
}

%%%%%%%%%%%%%%%%%%%%%%%%%%%%%%%%%%%%%%%%%%
%% optional
\appendixtitles{yes} %Leave argument "no" if all appendix headings stay EMPTY (then no dot is printed after "Appendix A"). If the appendix sections contain a heading then change the argument to "yes".
\appendixsections{multiple} %Leave argument "multiple" if there are multiple sections. Then a counter is printed ("Appendix A"). If there is only one appendix section then change the argument to "one" and no counter is printed ("Appendix").
\appendix
\section{Collision integral derivation}\label{Appendix1}
Here we will simplify the collision integral \eqref{eqn:collision_integral1}:
\begin{eqnarray}\label{eqn:collision_integral2}
	C_1[f] = \frac{(2\pi)^4}{2\mathrm E_1} \int \delta^4 (\sum_i P_{i}) | \mathrm M |^2 F[f] \prod_{k=2}^{4} \frac{\mathrm d^3 \vec{p}_k}{(2\pi)^3 2E_k}
\end{eqnarray}
Using the identity:
\begin{eqnarray}
	\delta^3 (\sum \vec p_i) = \int \exp{(i(\vec \lambda, \vec p_1 + \vec p_2 - \vec p_3 - \vec p_4))} \cdot \frac{\mathrm d^3 \vec \lambda}{(2\pi)^3},
\end{eqnarray}
and separating the angle integrations: 
\begin{eqnarray}
	\mathrm d\vec p_i = \mathrm d \varphi_i \mathrm d \cos\theta_i p_i^2 \mathrm d p_i = \varepsilon_i p_i \mathrm d \Omega_i \mathrm d \varepsilon_i
\end{eqnarray}
the integral takes the following form:
\begin{eqnarray}
	C_1[f] = \frac{\left|M\right|^2}{64\pi^3 \varepsilon_1} \int \delta (\varepsilon_1 + \varepsilon_2 - \varepsilon_3 - \varepsilon_4) \mathrm D F[f]
	\mathrm d \varepsilon_3 \mathrm d \varepsilon_4 \mathrm d \varepsilon_2, 
\end{eqnarray}
where D is defined as follows:
\begin{eqnarray}
	\mathrm D = \frac{p_2 p_3 p_4}{64\pi^5 } \int \lambda^2 \mathrm d \lambda \int e^{i(\vec p_1, \vec \lambda)} \mathrm d \Omega_{\lambda} \int e^{i(\vec p_2, \vec \lambda)} 
	\mathrm d \Omega_2  \int e^{i(\vec p_3, \vec \lambda)} \mathrm d \Omega_3 \int e^{i(\vec p_4, \vec \lambda)} \mathrm d \Omega_4 \,.
\end{eqnarray}
Taking into account that
\begin{align}
\begin{split}
	\int e^{i(\vec p_1, \vec \lambda)} \mathrm d \Omega_{\lambda} &= \int e^{i(p_1 \lambda \cos\theta_{\lambda})} \mathrm d \Omega_{\lambda}
	=\int_0^{2\pi} \mathrm d\varphi \int_{-1}^1 \mathrm d\cos\theta e^{i(p_1 \lambda \cos\theta_{\lambda})} = \\
	&=\frac{2\pi}{ip_1 \lambda} e^{ip_1 \lambda x} \big{|}^{x=1}_{x=-1}
	=\frac{2\pi}{p_1 \lambda} \frac{e^{ip_1 \lambda} - e^{-ip_1 \lambda}}{2i} \cdot 2 = \frac{4\pi}{p_1 \lambda} \sin(p_1 \lambda)
\end{split}
\end{align}
we can rewrite D as:
\begin{eqnarray}
	\mathrm D = \frac{4}{\pi p_1 } \int \frac{\mathrm d \lambda}{\lambda^2} \sin(p_1 \lambda) \sin(p_2 \lambda) \sin(p_3 \lambda) \sin(p_4 \lambda)
\end{eqnarray}
Using the Fourier transformation:
\begin{align}
\begin{split}
&-\sqrt{\frac{\pi}{2}} w \operatorname{Sign}(w) = \frac{1}{\sqrt{2 \pi}} \int^{\infty}_{-\infty} \frac{1}{x^2} e^{iwx} \mathrm d x = \\
&\frac{1}{\sqrt{2 \pi}} \left( \int^{\infty}_{0} \frac{1}{x^2} e^{iwx} \mathrm d x + \int^{\infty}_{0} \frac{1}{(-x)^2} e^{-iwx} \mathrm d x \right) = 
\frac{1}{\sqrt{2 \pi}}  \int^{\infty}_{0} \frac{1}{x^2} \left( e^{iwx} + e^{-iwx} \right) \mathrm d x,
\end{split}
\end{align}
we can simplify the integral in the formula for D:
\begin{align}\label{eq:intermidiate_d}
\begin{split}
	\mathrm D &= \frac{4}{\pi p_1 } \int^{\infty}_0 \frac{\mathrm d \lambda}{\lambda^2} \frac{e^{ip_1 \lambda} - e^{-ip_1 \lambda}}{2i} \frac{e^{ip_2 \lambda} - e^{-ip_2 \lambda}}{2i} \frac{e^{ip_3 \lambda} - e^{-ip_3 \lambda}}{2i} \frac{e^{ip_4 \lambda} - e^{-ip_4 \lambda}}{2i} =  \\
	&= \frac{1}{4\pi p_1 } \int^{\infty}_0 \frac{\mathrm d \lambda}{\lambda^2} \Big{(} (e^{i\lambda(p_1 + p_2)} - e^{i\lambda(p_2 - p_1)} - 
	-e^{i\lambda(p_1 - p_2)} + e^{i\lambda(-p_1 - p_2)})(e^{i\lambda(p_3 + p_4)} - \\
	&- e^{i\lambda(p_4 - p_3)} - e^{i\lambda(p_3 - p_4)} + e^{i\lambda(-p_3 - p_4)})\Big{)} =  \\
	&=\frac{1}{4\pi p_1 } \int^{\infty}_0 \frac{\mathrm d \lambda}{\lambda^2} \Big{(} e^{i\lambda(p_1 + p_2 + p_3 + p_4)} +
	%&\frac{1}{4\pi p_1 } \int^{\infty}_0 \frac{\mathrm d \lambda}{\lambda^2} \Big{(} e^{i\lambda(p_1 + p_2 + p_3 + p_4)} + ... \\
	 e^{-i\lambda(p_1 + p_2 + p_3 + p_4)} - e^{i\lambda(p_3 + p_4 + p_2 - p_1)} \\
	&-e^{-i\lambda(p_3 + p_4 + p_2 - p_1)} - e^{i\lambda(p_3 + p_4 + p_1 - p_2)} - 
	 e^{-i\lambda(p_3 + p_4 + p_1 - p_2)} + e^{i\lambda(p_3 + p_4 - p_1 - p_2)} + \\
	&+e^{-i\lambda(p_3 + p_4 - p_1 - p_2)} - e^{i\lambda(p_1 + p_2 + p_4 - p_3)} -
	 e^{-i\lambda(p_1 + p_2 + p_4 - p_3)} + e^{i\lambda(p_4 - p_3 + p_2 - p_1)} + \\
	&+e^{-i\lambda(p_4 - p_3 + p_2 - p_1)} + e^{i\lambda(p_4 - p_3 + p_1 - p_2)} + 
	 e^{-i\lambda(p_4 - p_3 + p_1 - p_2)} - e^{i\lambda(p_4 - p_3 - p_1 - p_2)}  \\
	&-e^{-i\lambda(p_4 - p_3 - p_1 - p_2)}\Big{)} = \\
	&= \frac{1}{4\pi p_1 }(-\pi) 
	%&...-e^{-i\lambda(p_4 - p_3 - p_1 - p_2)}\Big{)} = \frac{1}{4\pi p_1 } \times \\
	 (|p_1 + p_2 + p_3 + p_4| - |p_3 + p_4 + p_2 - p_1|
	-|p_3 + p_4 + p_1 - p_2|  \\
	&+ |p_3 + p_4 - p_1 - p_2| -|p_1 + p_2 + p_4 - p_3| + |p_4 - p_3 + p_2 - p_1| \\
	&+|p_4 - p_3 + p_1 - p_2| - |p_4 - p_3 - p_1 - p_2|) = \\
	&=\frac{1}{4\pi p_1 } (-\pi) (-4 \min \{ p_1,p_2,p_3,p_4\}) = \frac{\min \{ p_1,p_2,p_3,p_4\}}{p_1}
\end{split}
\end{align}
The last step in this equation \eqref{eq:intermidiate_d} (changing to the minimum function between the 4 momenta) can be easily done by checking one of the possibilities, - for example, 
the case when $p_1 = \min \{ p_1,p_2,p_3,p_4\}$ (or $p_1 < p_2 < p_3 < p_4$). Taking into account the 4-momentum conservation: $P_1 + P_2 = P_3 + P_4$, we get the final result:
\begin{eqnarray}\label{eqn:integ_1}
	C[f(\varepsilon_1)]  = \frac{\left|M\right|^2}{64\pi^3 \varepsilon_1} \int \int \mathrm d \varepsilon_3 \mathrm d \varepsilon_4 
	\mathrm D F[f]
\end{eqnarray}
where $ D = \frac{1}{p_1} \min\{p_1,p_2,p_3,p_4\} $.
In order to change the formula \eqref{eqn:integ_1} to the integration over momentum, we can use the connection between energy and momentum in the relativistic case: 
$\varepsilon^2 = p^2 + m^2 \rightarrow \varepsilon \mathrm d \varepsilon = p \mathrm d p$, so that the equation \eqref{eqn:integ_1} takes form:
\begin{eqnarray}\label{eqn:integ_2}
	C[f(p_1)]  = \frac{\left|M\right|^2}{64\pi^3 \varepsilon_1} \int \int \frac{p_3 p_4}{\varepsilon_3 \varepsilon_4} \mathrm d p_3 \mathrm d p_4 
	\mathrm D F[f] \,. 
\end{eqnarray}


%=====================================
% References, variant A: internal bibliography
%=====================================
%\reftitle{References}
%\begin{thebibliography}{999}
% Reference 1
%\bibitem[Author1(year)]{ref-journal}
%Author1, T. The title of the cited article. {\em Journal Abbreviation} {\bf 2008}, {\em 10}, 142-149, doi:xxxxx.
% Reference 2
%\bibitem[Author2(year)]{ref-book}
%Author2, L. The title of the cited contribution. In {\em The Book Title}; Editor1, F., Editor2, A., Eds.; Publishing House: City, Country, 2007; pp. 32-58, ISBN.
%\end{thebibliography}

% The following MDPI journals use author-date citation: Arts, Econometrics, Economies, Genealogy, Humanities, IJFS, JRFM, Laws, Religions, Risks, Social Sciences. For those journals, please follow the formatting guidelines on http://www.mdpi.com/authors/references
% To cite two works by the same author: \citeauthor{ref-journal-1a} (\citeyear{ref-journal-1a}, \citeyear{ref-journal-1b}). This produces: Whittaker (1967, 1975)
% To cite two works by the same author with specific pages: \citeauthor{ref-journal-3a} (\citeyear{ref-journal-3a}, p. 328; \citeyear{ref-journal-3b}, p.475). This produces: Wong (1999, p. 328; 2000, p. 475)

%=====================================
% References, variant B: external bibliography
%=====================================
\reftitle{References}
%\externalbibliography{yes}
\bibliography{references}

%%%%%%%%%%%%%%%%%%%%%%%%%%%%%%%%%%%%%%%%%%
%% optional
%\sampleavailability{Samples of the compounds ...... are available from the authors.}

%% for journal Sci
%\reviewreports{\\
%Reviewer 1 comments and authors’ response\\
%Reviewer 2 comments and authors’ response\\
%Reviewer 3 comments and authors’ response
%}

%%%%%%%%%%%%%%%%%%%%%%%%%%%%%%%%%%%%%%%%%%
\end{document}